\documentclass[11pt,a4paper]{article}

\usepackage{microtype}
\usepackage{xcolor}
\usepackage[margin=0.85in]{geometry} 
\usepackage{multirow}
\usepackage{amsmath,amsfonts,amssymb}
\usepackage{graphicx}
\usepackage[colorlinks=true, allcolors=blue]{hyperref}
\usepackage[utf8]{inputenc}
\usepackage[sc]{mathpazo}
\usepackage[T1]{fontenc}
\usepackage{parskip}
\title{Can a single image processing algorithm work equally well across all phases of DCE-MRI?}
\date{\vspace{-1.5ex}}
\author{Adam G. Tattersall$^{1,2}$, Keith A. Goatman$^2$, Lucy E. Kershaw$^1$,\\ Scott I. K. Semple$^1$, and Sonia Dahdouh$^2$}

\begin{document} 
\maketitle

\begin{abstract}
\noindent{}Image segmentation and registration are said to be challenging when applied to dynamic contrast enhanced MRI sequences (DCE-MRI). The contrast agent causes rapid changes in intensity in the region of interest and elsewhere, which can lead to false positive predictions for segmentation tasks and confound the image registration similarity metric. While it is widely assumed that contrast changes increase the difficulty of these tasks, to our knowledge no work has quantified these effects. 

In this paper we examine the effect of training with different ratios of contrast enhanced (CE) data on two popular tasks: segmentation with nnU-Net and Mask R-CNN and registration using VoxelMorph and VTN. We experimented further by strategically using the available datasets through pretraining and fine tuning with different splits of data. We found that to create a generalisable model, pretraining with CE data and fine tuning with non-CE data gave the best result. This interesting find could be expanded to other deep learning based image processing tasks with DCE-MRI and provide significant improvements to the models' performance.

\end{abstract}
\let\thefootnote\relax\footnotetext{$^1$University of Edinburgh, Edinburgh, UK. $^2$Canon Medical Research Europe, Edinburgh, UK.}
\let\thefootnote\relax\footnotetext{Send correspondence to Adam Tattersall: E-mail: adam.tattersall@ed.ac.uk}

\section{Introduction}
Dynamic Contrast Enhanced (DCE)-MRI examinations can be used to monitor microvascular perfusion by injecting the patient with a contrast agent and taking multiple rapid T1 weighted MR images over the course of $\sim$5 minutes.
A tracer kinetics model can be fitted to the intensity vs time data to gain estimates of microvascular parameters. \cite{Kershaw2006}, usually within a manually selected region of interest (ROI). This can be time consuming and any motion present can lead to imprecise results.
Image processing techniques, such as automated image segmentation and image registration, allow for faster and more accurate analysis. However, due to the rapidly changing intensities over the course of the dynamic series, application of these image processing techniques can be challenging \cite{Mahapatra2015}. Some pre-deep learning approaches dealt with these challenges by trying to reduce the temporal resolution using robust principal component analysis (PCA) to decompose the data into a sparse and low rank component \cite{Mahapatra2015}. Similarly, deep-learning based methods have used PCA to reduce the temporal dimension \cite{Warfield2018}.
Traditional and deep learning-based approaches both follow the premise that using a mix of images with and without contrast enhanced (CE) data leads to less accurate results. However, to the best of our knowledge, while this hypothesis was at the core of many deep-learning based methods, no one has attempted to quantify the impact of training with CE on the most common deep-learning based image processing tasks applied to DCE-MRI: namely image registration and segmentation.
In this work, different datasets and combinations of data with and without CE will be used to present a comprehensive view of the impact of training these popular techniques with CE data. 


\section{METHODS}
\subsection{Data}
Our experiments used two DCE-MRI datasets. The first contains 2D kidney DCE-MRI (K$\_$dataset) \cite{Lietzmann2012} from 13 patients with 375 images taken continuously at a temporal resolution of 1.6 seconds and a spatial resolution of (384, 348) for a total of 4875 images. Annotations for each image were drawn by an experienced clinical scientist (Lucy E.~Kershaw). The second contains 3D prostate DCE-MRI (P$\_$dataset) \cite{lemaitre} from 20 patients with 40 volumes acquired continuously at a temporal resolution of 6 seconds and a spatial resolution of (256, 192, 16). Annotations were made on T2 weighted MR images and registered to the DCE-MRI data\cite{lemaitre}.

Both datasets were initially split at a patient level with an 80:20 ratio. Each of these sets were then split further to create (1) a set containing images without CE (K$\_$NoCE and P$\_$NoCE), (2) a set containing images with CE (K$\_$CE and P$\_$CE) and (3) a set that was a mixture of both (K$\_$Mixed and P$\_$Mixed). Another set (4) was created containing a mix of images without CE and images just after CE injection until the highest intensity peak in the sequence (K$\_$Peak and P$\_$Peak). This contains the largest intensity change as the tissue enhances. After splitting according to CE, a high data imbalance appeared, favouring data with CE. To mitigate that concern, two different sampled datasets were created (K$\_$CE$\_$Sampled and P$\_$CE$\_$Sampled) and (K$\_$Mixed$\_$Sampled and P$\_$Mixed$\_$Sampled) by randomly sampling an equal number of images in each category, at a patient level, so that each dataset contained the same number of images. Five-fold cross validation was used in each case. We also explored the use of pre-training on either CE or non CE data for each image processing task and then fine-tuning on the remaining modality. This allowed us to explore other potential strategies using all the available data. 

\subsection{Image Segmentation}
\label{sec:title}

nnU-Net \cite{Isensee2021} is a self-configuring variation of U-Net which has achieved state-of-the-art performance across 23 datasets in international biomedical segmentation challenges \cite{Isensee2021}{.} It standardises pre-processing and parameter selection which allows for an easier comparison of results. Based on an encoder-decoder architecture, U-Net uses a contracting path to create feature maps, and an expansive path which performs pixelwise prediction by concatenating features maps from corresponding layers in the contracting and expansive paths.


Mask R-CNN \cite{He2017} is an instance segmentation model that extends Faster R-CNN \cite{Ren2017} by predicting a segmentation mask in parallel to object detection. It combines a ResNet backbone \cite{He2016} with a feature pyramid network \cite{yi2016} to create feature maps. These are passed into a region proposal network that outputs bounding boxes which are then classified and used for segmentation. Mask R-CNN achieved state-of-the-art performance by consistently ranking high in numerous instance segmentation challenges \cite{He2017}.


For consistency, both Mask-RCNN and nnU-Net used the pre-processing and augmentation that were deemed ideal by nnU-Net. For both datasets, the task was to predict binary segmentation masks for
both kidneys or whole prostate. Both approaches were used to assess the effect of training with CE on binary segmentation performance. Overall performance was evaluated using Dice score along with precision, recall and accuracy.

\subsection{Image Registration}

VoxelMorph \cite{Balakrishnan2019} is an unsupervised approach achieving state-of-the-art performance for biomedical image registration. It is based on a U-Net architecture and uses a loss function that captures image similarity and field smoothness. Mutual information (MI) was chosen as the similarity metric for VoxelMorph.


Volume Tweening Network (VTN)\cite{Zhao2019} is a registration method that consists of two different types of cascading registration subnetworks: affine and dense. The affine subnetwork, used first, aligns the input image with an affine transform using a basic convolutional network. The dense subnetwork uses an autoencoder which refines this output and produces a dense flow field of the same size as the input.

The choice of the reference image was the same for both approaches. When the dataset contained CE, the final image in the series was chosen for its similarity to the other images with CE. When no CE was present, the first image was arbitrarily chosen as reference. Dice score between the warped ground truth (GT) and the reference GT was used to evaluate registration accuracy.


\section{RESULTS}

Due to space constraints, results will only be shown for the sampled datasets. Results obtained using both sampled and non-sampled datasets followed the same trends.

\subsection{Image Segmentation}

Table \ref{tab:Segresultsunet} shows the results of the image segmentation tasks using nnU-Net. Overall, training with a mixture of images with and without CE gave the best results for each test set with respective Dice scores of 89.2 $\pm$ 4.3 and 81.4 $\pm$ 3.8 for K$\_$Mixed$\_$Sampled. For P$\_$Mixed$\_$Sampled, a Dice score of 91.1 $\pm$ 2.9 and 82.4 $\pm$ 3.7 was recorded. In contrast, training with only non-CE images (K$\_$NoCE or P$\_$NoCE) gave the worst results with Dice scores of 74.3 $\pm$ 3.8 and 78.2 $\pm$ 2.6 respectively when tested with CE images and 73.8 $\pm$ 3.9 and 81.4 $\pm$ 3.3 when tested using non-CE images.

When using a model pre-trained with non-CE images (K$\_$NoCE or P$\_$NoCE) and fine tuning with CE images (K$\_$CE$\_$Sampled or P$\_$CE$\_$Sampled), Dice scores of 88.5 $\pm$ 3.7 and 89.8 $\pm$ 2.4 when testing with CE images was achieved. A Dice score of 74.2 $\pm$ 4.8 and 77.9 $\pm$ 3.1 was obtained when testing with non-CE images. 

\begin{table}[ht]
\addtolength{\tabcolsep}{-2.5pt}
\scriptsize
\caption{Results from each of the experiments using nnU-Net for segmentation. The Dice score, precision, recall and accuracy were recorded.} 
\label{tab:Segresultsunet}
\begin{center}    
\begin{tabular}{|*{13}{c|}}  
\hline
\multicolumn{1}{|c}{} & \multicolumn{4}{|c|}{} & \multicolumn{8}{|c|}{Type of data used to train the model} \\ 
\cline{6-13}

\multicolumn{1}{|c}{} & \multicolumn{4}{|c}{} & \multicolumn{2}{|c}{} & 
\multicolumn{2}{|c}{} & \multicolumn{2}{|c}{} & \multicolumn{2}{|c|}{No CE} \\
\multicolumn{1}{|c}{Method} & \multicolumn{4}{|c}{Test Data} & \multicolumn{2}{|c}{Mixed} &
\multicolumn{2}{|c}{CE} & \multicolumn{2}{|c}{No CE} & \multicolumn{2}{|c|}{and early CE} \\
\multicolumn{1}{|c}{} & \multicolumn{4}{|c}{} & \multicolumn{2}{|c}{} &
\multicolumn{2}{|c}{} & \multicolumn{2}{|c}{} & \multicolumn{2}{|c|}{ uptake} \\
\cline{6-13} 

\multicolumn{1}{|c}{} & \multicolumn{4}{|c}{} & \multicolumn{1}{|c}{Kidney} & \multicolumn{1}{|c}{Prostate} &
\multicolumn{1}{|c}{Kidney} & \multicolumn{1}{|c}{Prostate} & \multicolumn{1}{|c}{Kidney} & \multicolumn{1}{|c}{Prostate} & \multicolumn{1}{|c}{Kidney} & \multicolumn{1}{|c|}{Prostate} \\\hline 

\multicolumn{1}{|c}{} & \multicolumn{2}{|c}{} & \multicolumn{2}{|c}{Dice $\pm$ std} & \multicolumn{1}{|c}{\textbf{89.2 $\pm$ 4.3}} & \multicolumn{1}{|c}{\textbf{91.1 $\pm$ 2.9}} &
\multicolumn{1}{|c}{88.7 $\pm$ 3.7} & \multicolumn{1}{|c}{90.2 $\pm$ 3.4} & \multicolumn{1}{|c}{74.3 $\pm$ 3.8} & \multicolumn{1}{|c}{78.2 $\pm$ 2.6} & \multicolumn{1}{|c}{78.7 $\pm$ 3.3} & \multicolumn{1}{|c|}{80.1 $\pm$ 2.9} \\
\cline{4-13}
\multicolumn{1}{|c}{} & \multicolumn{2}{|c}{CE} & \multicolumn{2}{|c}{Precision} & \multicolumn{1}{|c}{0.813} & \multicolumn{1}{|c}{0.904} &
\multicolumn{1}{|c}{0.784} & \multicolumn{1}{|c}{0.831} & \multicolumn{1}{|c}{0.616} & \multicolumn{1}{|c}{0.684} & \multicolumn{1}{|c}{0.724} & \multicolumn{1}{|c|}{0.712}\\
\cline{4-13}
\multicolumn{1}{|c}{nnU-Net} & \multicolumn{2}{|c}{} & \multicolumn{2}{|c}{Recall} & \multicolumn{1}{|c}{0.942} & \multicolumn{1}{|c}{0.948} &
\multicolumn{1}{|c}{0.938} & \multicolumn{1}{|c}{0.936} & \multicolumn{1}{|c}{0.931} & \multicolumn{1}{|c}{0.924} & \multicolumn{1}{|c}{0.926} & \multicolumn{1}{|c|}{0.931}\\ 
\cline{4-13}
\multicolumn{1}{|c}{(no} & \multicolumn{2}{|c}{ } & \multicolumn{2}{|c}{Accuracy} & \multicolumn{1}{|c}{0.987} & \multicolumn{1}{|c}{0.989} &
\multicolumn{1}{|c}{0.985} & \multicolumn{1}{|c}{0.988} & \multicolumn{1}{|c}{0.967} & \multicolumn{1}{|c}{0.976} & \multicolumn{1}{|c}{0.973} & \multicolumn{1}{|c|}{0.979}\\

\cline{2-13}

\multicolumn{1}{|c}{pretraining)} & \multicolumn{2}{|c}{} & \multicolumn{2}{|c}{Dice $\pm$ std} & \multicolumn{1}{|c}{81.4 $\pm$ 3.8} & \multicolumn{1}{|c}{82.4 $\pm$ 3.7} &
\multicolumn{1}{|c}{69.2 $\pm$ 4.3} & \multicolumn{1}{|c}{71.4 $\pm$ 3.4} & \multicolumn{1}{|c}{73.8 $\pm$ 3.9} & \multicolumn{1}{|c}{81.4 $\pm$ 3.3} & \multicolumn{1}{|c}{74.1 $\pm$ 3.4} & \multicolumn{1}{|c|}{81.8 $\pm$ 2.4} \\
\cline{4-13}
\multicolumn{1}{|c}{} & \multicolumn{2}{|c}{No CE} & \multicolumn{2}{|c}{Precision} & \multicolumn{1}{|c}{0.841} & \multicolumn{1}{|c}{0.855} &
\multicolumn{1}{|c}{0.611} & \multicolumn{1}{|c}{0.627} & \multicolumn{1}{|c}{0.639} & \multicolumn{1}{|c}{0.659} & \multicolumn{1}{|c}{0.682} & \multicolumn{1}{|c|}{0.682}\\
\cline{4-13}
\multicolumn{1}{|c}{} & \multicolumn{2}{|c}{} & \multicolumn{2}{|c}{Recall } & \multicolumn{1}{|c}{0.704} & \multicolumn{1}{|c}{0.729} &
\multicolumn{1}{|c}{0.713} & \multicolumn{1}{|c}{0.765} & \multicolumn{1}{|c}{0.872} & \multicolumn{1}{|c}{0.895} & \multicolumn{1}{|c}{0.825} & \multicolumn{1}{|c|}{0.887}\\ 
\cline{4-13}
\multicolumn{1}{|c}{} & \multicolumn{2}{|c}{ } & \multicolumn{2}{|c}{Accuracy} & \multicolumn{1}{|c}{0.969} & \multicolumn{1}{|c}{0.972} &
\multicolumn{1}{|c}{0.955} & \multicolumn{1}{|c}{0.962} & \multicolumn{1}{|c}{0.968} & \multicolumn{1}{|c}{0.971} & \multicolumn{1}{|c}{0.971} & \multicolumn{1}{|c|}{0.974}\\

\hline
\hline

\multicolumn{1}{|c}{} & \multicolumn{2}{|c}{} & \multicolumn{2}{|c}{Dice $\pm$ std} & \multicolumn{1}{|c}{-} & \multicolumn{1}{|c}{-} &
\multicolumn{1}{|c}{88.5 $\pm$ 3.7} & \multicolumn{1}{|c}{89.8 $\pm$ 2.4} & \multicolumn{1}{|c}{-} & \multicolumn{1}{|c}{-} & \multicolumn{1}{|c}{-} & \multicolumn{1}{|c|}{-} \\
\cline{4-13}
\multicolumn{1}{|c}{} & \multicolumn{2}{|c}{CE} & \multicolumn{2}{|c}{Precision } & \multicolumn{1}{|c}{-} & \multicolumn{1}{|c}{-} &
\multicolumn{1}{|c}{0.843} & \multicolumn{1}{|c}{0.856} & \multicolumn{1}{|c}{-} & \multicolumn{1}{|c}{-} & \multicolumn{1}{|c}{-} & \multicolumn{1}{|c|}{-}\\
\cline{4-13}
\multicolumn{1}{|c}{nnU-Net} & \multicolumn{2}{|c}{ } & \multicolumn{2}{|c}{Recall} & \multicolumn{1}{|c}{-} & \multicolumn{1}{|c}{-} &
\multicolumn{1}{|c}{0.924} & \multicolumn{1}{|c}{0.926} & \multicolumn{1}{|c}{-} & \multicolumn{1}{|c}{-} & \multicolumn{1}{|c}{-} & \multicolumn{1}{|c|}{-}\\ 
\cline{4-13}
\multicolumn{1}{|c}{(pretrained} & \multicolumn{2}{|c}{} & \multicolumn{2}{|c}{Accuracy} & \multicolumn{1}{|c}{-} & \multicolumn{1}{|c}{-} &
\multicolumn{1}{|c}{0.986} & \multicolumn{1}{|c}{0.989} & \multicolumn{1}{|c}{-} & \multicolumn{1}{|c}{-} & \multicolumn{1}{|c}{-} & \multicolumn{1}{|c|}{-}\\

\cline{2-13}

\multicolumn{1}{|c}{with no} & \multicolumn{2}{|c}{} & \multicolumn{2}{|c}{Dice $\pm$ std} & \multicolumn{1}{|c}{-} & \multicolumn{1}{|c}{-} &
\multicolumn{1}{|c}{74.2 $\pm$ 4.8} & \multicolumn{1}{|c}{77.9 $\pm$ 3.1} & \multicolumn{1}{|c}{-} & \multicolumn{1}{|c}{-} & \multicolumn{1}{|c}{-} & \multicolumn{1}{|c|}{-} \\
\cline{4-13}
\multicolumn{1}{|c}{CE)} & \multicolumn{2}{|c}{No CE} & \multicolumn{2}{|c}{Precision} & \multicolumn{1}{|c}{-} & \multicolumn{1}{|c}{-} &
\multicolumn{1}{|c}{0.725} & \multicolumn{1}{|c}{0.769} & \multicolumn{1}{|c}{-} & \multicolumn{1}{|c}{-} & \multicolumn{1}{|c}{-} & \multicolumn{1}{|c|}{-}\\
\cline{4-13}
\multicolumn{1}{|c}{} & \multicolumn{2}{|c}{} & \multicolumn{2}{|c}{Recall} & \multicolumn{1}{|c}{-} & \multicolumn{1}{|c}{-} &
\multicolumn{1}{|c}{0.776} & \multicolumn{1}{|c}{0.781} & \multicolumn{1}{|c}{-} & \multicolumn{1}{|c}{-} & \multicolumn{1}{|c}{-} & \multicolumn{1}{|c|}{-}\\ 
\cline{4-13}
\multicolumn{1}{|c}{} & \multicolumn{2}{|c}{ } & \multicolumn{2}{|c}{Accuracy} & \multicolumn{1}{|c}{-} & \multicolumn{1}{|c}{-} &
\multicolumn{1}{|c}{0.971} & \multicolumn{1}{|c}{0.978} & \multicolumn{1}{|c}{-} & \multicolumn{1}{|c}{-} & \multicolumn{1}{|c}{-} & \multicolumn{1}{|c|}{-}\\

\hline
\hline

\multicolumn{1}{|c}{} & \multicolumn{2}{|c}{} & \multicolumn{2}{|c}{Dice $\pm$ std} & \multicolumn{1}{|c}{-} & \multicolumn{1}{|c}{-} &
\multicolumn{1}{|c}{-} & \multicolumn{1}{|c}{-} & \multicolumn{1}{|c}{87.4 $\pm$ 2.8} & \multicolumn{1}{|c}{88.4 $\pm$ 2.6} & \multicolumn{1}{|c}{-} & \multicolumn{1}{|c|}{-} \\
\cline{4-13}
\multicolumn{1}{|c}{} & \multicolumn{2}{|c}{CE} & \multicolumn{2}{|c}{Precision} & \multicolumn{1}{|c}{-} & \multicolumn{1}{|c}{-} &
\multicolumn{1}{|c}{-} & \multicolumn{1}{|c}{-} & \multicolumn{1}{|c}{0.821} & \multicolumn{1}{|c}{0.836} & \multicolumn{1}{|c}{-} & \multicolumn{1}{|c|}{-}\\
\cline{4-13}
\multicolumn{1}{|c}{nnU-Net} & \multicolumn{2}{|c}{ } & \multicolumn{2}{|c}{Recall} & \multicolumn{1}{|c}{-} & \multicolumn{1}{|c}{-} &
\multicolumn{1}{|c}{-} & \multicolumn{1}{|c}{-} & \multicolumn{1}{|c}{0.898} & \multicolumn{1}{|c}{0.902} & \multicolumn{1}{|c}{-} & \multicolumn{1}{|c|}{-}\\ 
\cline{4-13}
\multicolumn{1}{|c}{(pretrained} & \multicolumn{2}{|c}{ } & \multicolumn{2}{|c}{Accuracy} & \multicolumn{1}{|c}{-} & \multicolumn{1}{|c}{-} &
\multicolumn{1}{|c}{-} & \multicolumn{1}{|c}{-} & \multicolumn{1}{|c}{0.981} & \multicolumn{1}{|c}{0.983} & \multicolumn{1}{|c}{-} & \multicolumn{1}{|c|}{-}\\

\cline{2-13}

\multicolumn{1}{|c}{with CE)} & \multicolumn{2}{|c}{} & \multicolumn{2}{|c}{Dice $\pm$ std} & \multicolumn{1}{|c}{-} & \multicolumn{1}{|c}{-} &
\multicolumn{1}{|c}{-} & \multicolumn{1}{|c}{-} & \multicolumn{1}{|c}{\textbf{83.1 $\pm$ 2.6}} & \multicolumn{1}{|c}{\textbf{85.7 $\pm$ 3.2}} & \multicolumn{1}{|c}{-} & \multicolumn{1}{|c|}{-} \\
\cline{4-13}
\multicolumn{1}{|c}{} & \multicolumn{2}{|c}{No CE} & \multicolumn{2}{|c}{Precision} & \multicolumn{1}{|c}{-} & \multicolumn{1}{|c}{-} &
\multicolumn{1}{|c}{-} & \multicolumn{1}{|c}{-} & \multicolumn{1}{|c}{0.819} & \multicolumn{1}{|c}{0.864} & \multicolumn{1}{|c}{-} & \multicolumn{1}{|c|}{-}\\
\cline{4-13}
\multicolumn{1}{|c}{} & \multicolumn{2}{|c}{} & \multicolumn{2}{|c}{Recall} & \multicolumn{1}{|c}{-} & \multicolumn{1}{|c}{-} &
\multicolumn{1}{|c}{-} & \multicolumn{1}{|c}{-} & \multicolumn{1}{|c}{0.868} & \multicolumn{1}{|c}{0.882} & \multicolumn{1}{|c}{-} & \multicolumn{1}{|c|}{-}\\ 
\cline{4-13}
\multicolumn{1}{|c}{} & \multicolumn{2}{|c}{ } & \multicolumn{2}{|c}{Accuracy} & \multicolumn{1}{|c}{-} & \multicolumn{1}{|c}{-} &
\multicolumn{1}{|c}{-} & \multicolumn{1}{|c}{-} & \multicolumn{1}{|c}{0.972} & \multicolumn{1}{|c}{0.974} & \multicolumn{1}{|c}{-} & \multicolumn{1}{|c|}{-}\\
\hline

\end{tabular}
\end{center}
\end{table}

\begin{figure*}[!htb]
\centering
\begin{minipage}[b]{.19\textwidth}
  \centering
  \centerline{\includegraphics[width=3.2cm]{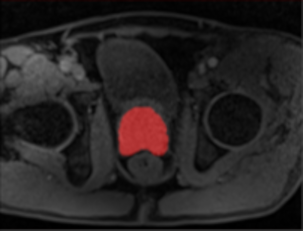}}
  \medskip
\end{minipage}
\begin{minipage}[b]{.19\textwidth}
  \centering
  \centerline{\includegraphics[width=3.2cm]{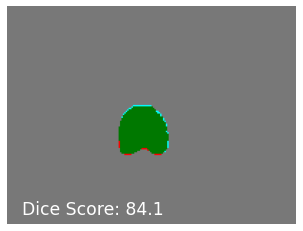}}
  \medskip
\end{minipage}
\begin{minipage}[b]{.19\textwidth}
  \centering
  \centerline{\includegraphics[width=3.2cm]{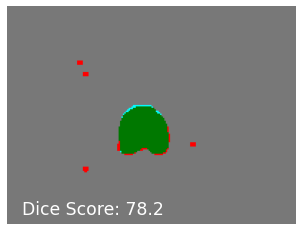}}
  \medskip
\end{minipage}
\begin{minipage}[b]{.19\textwidth}
  \centering
  \centerline{\includegraphics[width=3.2cm]{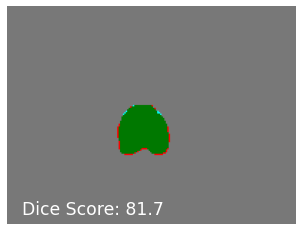}}
  \medskip
\end{minipage}
\begin{minipage}[b]{.19\textwidth}
  \centering
  \centerline{\includegraphics[width=3.2cm]{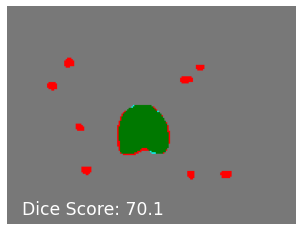}}
  \medskip
\end{minipage}

{\footnotesize
\begin{minipage}[b]{.19\textwidth}
  \centering
  \centerline{\includegraphics[width=3.2cm]{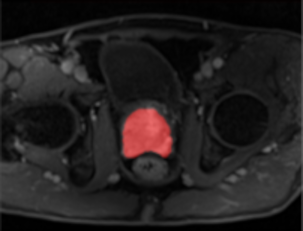}}
  \centerline{(a) GT Image and Mask}\medskip
\end{minipage}
\begin{minipage}[b]{.19\textwidth}
  \centering
  \centerline{\includegraphics[width=3.2cm]{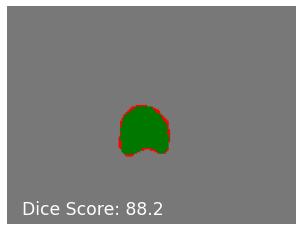}}
  \centerline{(b) Pretrained CE}\medskip
\end{minipage}
\begin{minipage}[b]{.19\textwidth}
  \centering
  \centerline{\includegraphics[width=3.2cm]{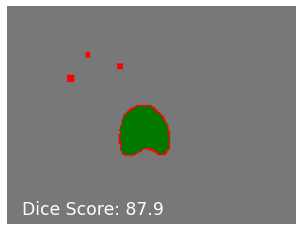}}
  \centerline{(c) Pretrained No CE}\medskip
\end{minipage}
\begin{minipage}[b]{.19\textwidth}
  \centering
  \centerline{\includegraphics[width=3.2cm]{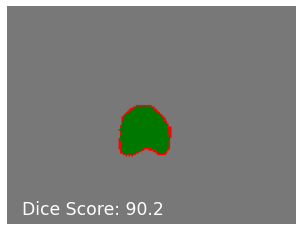}}
  \centerline{(d) Mixed}\medskip
\end{minipage}
\begin{minipage}[b]{.19\textwidth}
  \centering
  \centerline{\includegraphics[width=3.2cm]{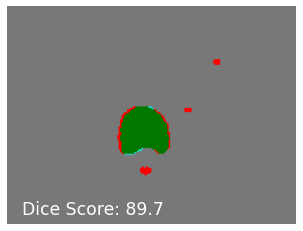}}
  \centerline{(d) CE}\medskip
\end{minipage}
}
\caption{nnU-Net segmentation results on 3D prostate data. (a) ground truth image (top row is no CE and bottom row is CE) with overlayed mask, (b) model pretrained using CE and refined on non CE data, (c) model pretrained using non CE and refined on CE data, (d) model trained using mixed data, (e) model trained using CE data. Green shows true positives, red shows false positives and blue shows false negatives.}
\label{fig:unet_prost}
\bigskip
\bigskip
\end{figure*}

Figure \ref{fig:unet_prost} shows some results for models trained using nnU-Net. Models pretrained with no CE data then fine tuned with CE data (c) and models trained with CE data (d) both showed examples where the model made false postive predictions. 

\bigskip

\begin{figure*}[!htb]
\centering
\begin{minipage}[b]{.19\textwidth}
  \centering
  \centerline{\includegraphics[width=3.2cm]{images/gt/noceimage.png}}
  \medskip
\end{minipage}
\begin{minipage}[b]{.19\textwidth}
  \centering
  \centerline{\includegraphics[width=3.2cm]{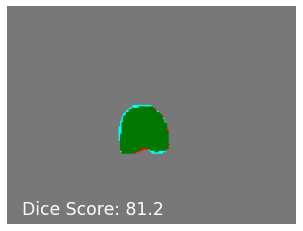}}
  \medskip
\end{minipage}
\begin{minipage}[b]{.19\textwidth}
  \centering
  \centerline{\includegraphics[width=3.2cm]{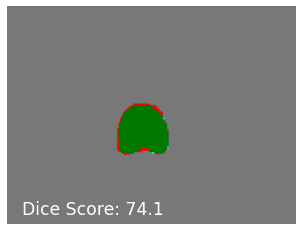}}
  \medskip
\end{minipage}
\begin{minipage}[b]{.19\textwidth}
  \centering
  \centerline{\includegraphics[width=3.2cm]{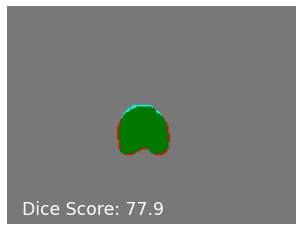}}
  \medskip
\end{minipage}
\begin{minipage}[b]{.19\textwidth}
  \centering
  \centerline{\includegraphics[width=3.2cm]{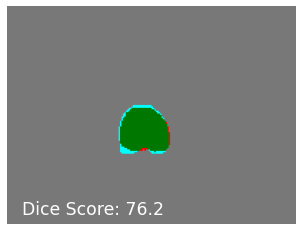}}
  \medskip
\end{minipage}

{\footnotesize
\begin{minipage}[b]{.19\textwidth}
  \centering
  \centerline{\includegraphics[width=3.2cm]{images/gt/ceimage.png}}
  \centerline{(a) GT Image and Mask}\medskip
\end{minipage}
\begin{minipage}[b]{.19\textwidth}
  \centering
  \centerline{\includegraphics[width=3.2cm]{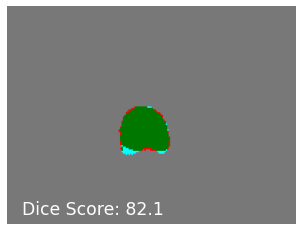}}
  \centerline{(b) Pretrained CE}\medskip
\end{minipage}
\begin{minipage}[b]{.19\textwidth}
  \centering
  \centerline{\includegraphics[width=3.2cm]{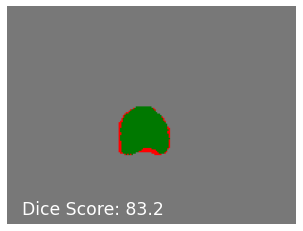}}
  \centerline{(c) Pretrained No CE}\medskip
\end{minipage}
\begin{minipage}[b]{.19\textwidth}
  \centering
  \centerline{\includegraphics[width=3.2cm]{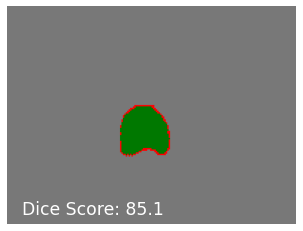}}
  \centerline{(d) Mixed}\medskip
\end{minipage}
\begin{minipage}[b]{.19\textwidth}
  \centering
  \centerline{\includegraphics[width=3.2cm]{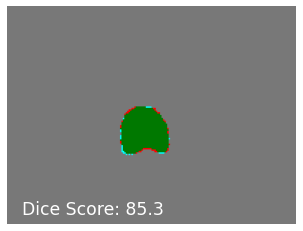}}
  \centerline{(e) CE}\medskip
\end{minipage}
}
\caption{Mask R-CNN segmentation results on 3D prostate data. (a) ground truth image (top row is no CE and bottom row is CE) with overlayed mask, (b) model pretrained using CE and refined on non CE data, (c) model pretrained using non CE and refined on CE data, (d) model trained using mixed data, (e) model trained using CE data. Green shows true positives, red shows false positives and blue shows false negatives.}
\label{fig:mask_prost}
\end{figure*}


\begin{table}[ht]
\addtolength{\tabcolsep}{-2.5pt}
\scriptsize
\caption{Results from each of the experiments using Mask R-CNN for segmentation. The Dice score, precision, recall and accuracy was recorded.} 
\label{tab:Segresultsmask}
\begin{center}    
\begin{tabular}{|*{13}{c|}}  
\hline
\multicolumn{1}{|c}{} & \multicolumn{4}{|c|}{} & \multicolumn{8}{|c|}{Type of data used to train the model} \\ 
\cline{6-13}

\multicolumn{1}{|c}{} & \multicolumn{4}{|c}{} & \multicolumn{2}{|c}{} & 
\multicolumn{2}{|c}{} & \multicolumn{2}{|c}{} & \multicolumn{2}{|c|}{No CE} \\
\multicolumn{1}{|c}{Method} & \multicolumn{4}{|c}{Test Data} & \multicolumn{2}{|c}{Mixed} &
\multicolumn{2}{|c}{CE} & \multicolumn{2}{|c}{No CE} & \multicolumn{2}{|c|}{and early CE} \\
\multicolumn{1}{|c}{} & \multicolumn{4}{|c}{} & \multicolumn{2}{|c}{} &
\multicolumn{2}{|c}{} & \multicolumn{2}{|c}{} & \multicolumn{2}{|c|}{ uptake} \\
\cline{6-13} 

\multicolumn{1}{|c}{} & \multicolumn{4}{|c}{} & \multicolumn{1}{|c}{Kidney} & \multicolumn{1}{|c}{Prostate} &
\multicolumn{1}{|c}{Kidney} & \multicolumn{1}{|c}{Prostate} & \multicolumn{1}{|c}{Kidney} & \multicolumn{1}{|c}{Prostate} & \multicolumn{1}{|c}{Kidney} & \multicolumn{1}{|c|}{Prostate} \\\hline 

\multicolumn{1}{|c}{} & \multicolumn{2}{|c}{} & \multicolumn{2}{|c}{Dice $\pm$ std} & \multicolumn{1}{|c}{\textbf{86.3 $\pm$ 3.6}} & \multicolumn{1}{|c}{\textbf{85.7 $\pm$ 3.1}} &
\multicolumn{1}{|c}{85.4 $\pm$ 3.5} & \multicolumn{1}{|c}{85.1 $\pm$ 4.7} & \multicolumn{1}{|c}{72.9 $\pm$ 3.3} & \multicolumn{1}{|c}{71.3 $\pm$ 3.1} & \multicolumn{1}{|c}{73.7 $\pm$ 2.8} & \multicolumn{1}{|c|}{72.4 $\pm$ 2.3} \\
\cline{4-13} 
\multicolumn{1}{|c}{} & \multicolumn{2}{|c}{CE} & \multicolumn{2}{|c}{Precision} & \multicolumn{1}{|c}{0.802} & \multicolumn{1}{|c}{0.798} &
\multicolumn{1}{|c}{0.773} & \multicolumn{1}{|c}{0.767} & \multicolumn{1}{|c}{0.614} & \multicolumn{1}{|c}{0.608} & \multicolumn{1}{|c}{0.714} & \multicolumn{1}{|c|}{0.716}\\
\cline{4-13}
\multicolumn{1}{|c}{Mask R-CNN} & \multicolumn{2}{|c}{} & \multicolumn{2}{|c}{Recall} & \multicolumn{1}{|c}{0.937} & \multicolumn{1}{|c}{0.924} &
\multicolumn{1}{|c}{0.921} & \multicolumn{1}{|c}{0.914} & \multicolumn{1}{|c}{0.927} & \multicolumn{1}{|c}{0.923} & \multicolumn{1}{|c}{0.932} & \multicolumn{1}{|c|}{0.929}\\ 
\cline{4-13}
\multicolumn{1}{|c}{(no} & \multicolumn{2}{|c}{ } & \multicolumn{2}{|c}{Accuracy} & \multicolumn{1}{|c}{0.981} & \multicolumn{1}{|c}{0.974} &
\multicolumn{1}{|c}{0.972} & \multicolumn{1}{|c}{0.967} & \multicolumn{1}{|c}{0.959} & \multicolumn{1}{|c}{0.953} & \multicolumn{1}{|c}{0.964} & \multicolumn{1}{|c|}{0.961}\\

\cline{2-13}

\multicolumn{1}{|c}{pretraining)} & \multicolumn{2}{|c}{} & \multicolumn{2}{|c}{Dice $\pm$ std} & \multicolumn{1}{|c}{79.2 $\pm$ 3.9} & \multicolumn{1}{|c}{78.1 $\pm$ 3.3} &
\multicolumn{1}{|c}{76.4 $\pm$ 2.9} & \multicolumn{1}{|c}{74.8 $\pm$ 3.2} & \multicolumn{1}{|c}{73.4 $\pm$ 2.9} & \multicolumn{1}{|c}{72.4 $\pm$ 3.1} & \multicolumn{1}{|c}{74.6 $\pm$ 2.1} & \multicolumn{1}{|c|}{73.2 $\pm$ 2.9} \\
\cline{4-13}
\multicolumn{1}{|c}{} & \multicolumn{2}{|c}{No CE} & \multicolumn{2}{|c}{Precision} & \multicolumn{1}{|c}{0.829} & \multicolumn{1}{|c}{0.826} &
\multicolumn{1}{|c}{0.621} & \multicolumn{1}{|c}{0.619} & \multicolumn{1}{|c}{0.617} & \multicolumn{1}{|c}{0.614} & \multicolumn{1}{|c}{0.620} & \multicolumn{1}{|c|}{0.618}\\
\cline{4-13}
\multicolumn{1}{|c}{} & \multicolumn{2}{|c}{} & \multicolumn{2}{|c}{Recall } & \multicolumn{1}{|c}{0.714} & \multicolumn{1}{|c}{0.712} &
\multicolumn{1}{|c}{0.719} & \multicolumn{1}{|c}{0.718} & \multicolumn{1}{|c}{0.715} & \multicolumn{1}{|c}{0.712} & \multicolumn{1}{|c}{0.718} & \multicolumn{1}{|c|}{0.717}\\ 
\cline{4-13}
\multicolumn{1}{|c}{} & \multicolumn{2}{|c}{ } & \multicolumn{2}{|c}{Accuracy} & \multicolumn{1}{|c}{0.964} & \multicolumn{1}{|c}{0.961} &
\multicolumn{1}{|c}{0.958} & \multicolumn{1}{|c}{0.956} & \multicolumn{1}{|c}{0.954} & \multicolumn{1}{|c}{0.951} & \multicolumn{1}{|c}{0.957} & \multicolumn{1}{|c|}{0.956}\\

\hline
\hline

\multicolumn{1}{|c}{} & \multicolumn{2}{|c}{} & \multicolumn{2}{|c}{Dice $\pm$ std} & \multicolumn{1}{|c}{-} & \multicolumn{1}{|c}{-} &
\multicolumn{1}{|c}{84.7 $\pm$ 3.2} & \multicolumn{1}{|c}{84.5 $\pm$ 3.4} & \multicolumn{1}{|c}{-} & \multicolumn{1}{|c}{-} & \multicolumn{1}{|c}{-} & \multicolumn{1}{|c|}{-} \\
\cline{4-13}
\multicolumn{1}{|c}{} & \multicolumn{2}{|c}{CE} & \multicolumn{2}{|c}{Precision } & \multicolumn{1}{|c}{-} & \multicolumn{1}{|c}{-} &
\multicolumn{1}{|c}{0.794} & \multicolumn{1}{|c}{0.791} & \multicolumn{1}{|c}{-} & \multicolumn{1}{|c}{-} & \multicolumn{1}{|c}{-} & \multicolumn{1}{|c|}{-}\\
\cline{4-13}
\multicolumn{1}{|c}{Mask R-CNN} & \multicolumn{2}{|c}{ } & \multicolumn{2}{|c}{Recall} & \multicolumn{1}{|c}{-} & \multicolumn{1}{|c}{-} &
\multicolumn{1}{|c}{0.921} & \multicolumn{1}{|c}{0.918} & \multicolumn{1}{|c}{-} & \multicolumn{1}{|c}{-} & \multicolumn{1}{|c}{-} & \multicolumn{1}{|c|}{-}\\ 
\cline{4-13}
\multicolumn{1}{|c}{(pretrained} & \multicolumn{2}{|c}{} & \multicolumn{2}{|c}{Accuracy} & \multicolumn{1}{|c}{-} & \multicolumn{1}{|c}{-} &
\multicolumn{1}{|c}{0.978} & \multicolumn{1}{|c}{0.969} & \multicolumn{1}{|c}{-} & \multicolumn{1}{|c}{-} & \multicolumn{1}{|c}{-} & \multicolumn{1}{|c|}{-}\\

\cline{2-13}

\multicolumn{1}{|c}{with no} & \multicolumn{2}{|c}{} & \multicolumn{2}{|c}{Dice $\pm$ std} & \multicolumn{1}{|c}{-} & \multicolumn{1}{|c}{-} &
\multicolumn{1}{|c}{74.1 $\pm$ 3.9} & \multicolumn{1}{|c}{73.8 $\pm$ 3.4} & \multicolumn{1}{|c}{-} & \multicolumn{1}{|c}{-} & \multicolumn{1}{|c}{-} & \multicolumn{1}{|c|}{-} \\
\cline{4-13}
\multicolumn{1}{|c}{CE)} & \multicolumn{2}{|c}{No CE} & \multicolumn{2}{|c}{Precision} & \multicolumn{1}{|c}{-} & \multicolumn{1}{|c}{-} &
\multicolumn{1}{|c}{0.649} & \multicolumn{1}{|c}{0.646} & \multicolumn{1}{|c}{-} & \multicolumn{1}{|c}{-} & \multicolumn{1}{|c}{-} & \multicolumn{1}{|c|}{-}\\
\cline{4-13}
\multicolumn{1}{|c}{} & \multicolumn{2}{|c}{} & \multicolumn{2}{|c}{Recall} & \multicolumn{1}{|c}{-} & \multicolumn{1}{|c}{-} &
\multicolumn{1}{|c}{0.728} & \multicolumn{1}{|c}{0.724} & \multicolumn{1}{|c}{-} & \multicolumn{1}{|c}{-} & \multicolumn{1}{|c}{-} & \multicolumn{1}{|c|}{-}\\ 
\cline{4-13}
\multicolumn{1}{|c}{} & \multicolumn{2}{|c}{ } & \multicolumn{2}{|c}{Accuracy} & \multicolumn{1}{|c}{-} & \multicolumn{1}{|c}{-} &
\multicolumn{1}{|c}{0.956} & \multicolumn{1}{|c}{0.953} & \multicolumn{1}{|c}{-} & \multicolumn{1}{|c}{-} & \multicolumn{1}{|c}{-} & \multicolumn{1}{|c|}{-}\\

\hline
\hline

\multicolumn{1}{|c}{} & \multicolumn{2}{|c}{} & \multicolumn{2}{|c}{Dice $\pm$ std} & \multicolumn{1}{|c}{-} & \multicolumn{1}{|c}{-} &
\multicolumn{1}{|c}{-} & \multicolumn{1}{|c}{-} & \multicolumn{1}{|c}{85.1 $\pm$ 2.1} & \multicolumn{1}{|c}{84.9 $\pm$ 3.7} & \multicolumn{1}{|c}{-} & \multicolumn{1}{|c|}{-} \\
\cline{4-13}
\multicolumn{1}{|c}{} & \multicolumn{2}{|c}{CE} & \multicolumn{2}{|c}{Precision} & \multicolumn{1}{|c}{-} & \multicolumn{1}{|c}{-} &
\multicolumn{1}{|c}{-} & \multicolumn{1}{|c}{-} & \multicolumn{1}{|c}{0.799} & \multicolumn{1}{|c}{0.791} & \multicolumn{1}{|c}{-} & \multicolumn{1}{|c|}{-}\\
\cline{4-13}
\multicolumn{1}{|c}{Mask R-CNN} & \multicolumn{2}{|c}{ } & \multicolumn{2}{|c}{Recall} & \multicolumn{1}{|c}{-} & \multicolumn{1}{|c}{-} &
\multicolumn{1}{|c}{-} & \multicolumn{1}{|c}{-} & \multicolumn{1}{|c}{0.921} & \multicolumn{1}{|c}{0.914} & \multicolumn{1}{|c}{-} & \multicolumn{1}{|c|}{-}\\ 
\cline{4-13}
\multicolumn{1}{|c}{(pretrained} & \multicolumn{2}{|c}{ } & \multicolumn{2}{|c}{Accuracy} & \multicolumn{1}{|c}{-} & \multicolumn{1}{|c}{-} &
\multicolumn{1}{|c}{-} & \multicolumn{1}{|c}{-} & \multicolumn{1}{|c}{0.978} & \multicolumn{1}{|c}{0.971} & \multicolumn{1}{|c}{-} & \multicolumn{1}{|c|}{-}\\

\cline{2-13}

\multicolumn{1}{|c}{with CE)} & \multicolumn{2}{|c}{} & \multicolumn{2}{|c}{Dice $\pm$ std} & \multicolumn{1}{|c}{-} & \multicolumn{1}{|c}{-} &
\multicolumn{1}{|c}{-} & \multicolumn{1}{|c}{-} & \multicolumn{1}{|c}{\textbf{81.3 $\pm$ 2.4}} & \multicolumn{1}{|c}{\textbf{80.8 $\pm$ 3.1}} & \multicolumn{1}{|c}{-} & \multicolumn{1}{|c|}{-} \\
\cline{4-13}
\multicolumn{1}{|c}{} & \multicolumn{2}{|c}{No CE} & \multicolumn{2}{|c}{Precision} & \multicolumn{1}{|c}{-} & \multicolumn{1}{|c}{-} &
\multicolumn{1}{|c}{-} & \multicolumn{1}{|c}{-} & \multicolumn{1}{|c}{0.831} & \multicolumn{1}{|c}{0.827} & \multicolumn{1}{|c}{-} & \multicolumn{1}{|c|}{-}\\
\cline{4-13}
\multicolumn{1}{|c}{} & \multicolumn{2}{|c}{} & \multicolumn{2}{|c}{Recall} & \multicolumn{1}{|c}{-} & \multicolumn{1}{|c}{-} &
\multicolumn{1}{|c}{-} & \multicolumn{1}{|c}{-} & \multicolumn{1}{|c}{0.718} & \multicolumn{1}{|c}{0.716} & \multicolumn{1}{|c}{-} & \multicolumn{1}{|c|}{-}\\ 
\cline{4-13}
\multicolumn{1}{|c}{} & \multicolumn{2}{|c}{ } & \multicolumn{2}{|c}{Accuracy} & \multicolumn{1}{|c}{-} & \multicolumn{1}{|c}{-} &
\multicolumn{1}{|c}{-} & \multicolumn{1}{|c}{-} & \multicolumn{1}{|c}{0.971} & \multicolumn{1}{|c}{0.968} & \multicolumn{1}{|c}{-} & \multicolumn{1}{|c|}{-}\\
\hline

\end{tabular}
\end{center}
\end{table}

Table \ref{tab:Segresultsmask} shows the results of the image segmentation task using Mask R-CNN. Overall, training with a mixture of images with and without CE gave the best results for each test set with respective Dice scores of 86.3 $\pm$ 3.6 and 79.2 $\pm$ 3.9 for K$\_$Mixed$\_$Sampled. For P$\_$Mixed$\_$Sampled, a Dice score of 85.7 $\pm$ 3.1 and 78.1 $\pm$ 3.3 were recorded. In contrast, training with only images without CE (K$\_$NoCE or P$\_$NoCE) gave the worst results with respective Dice scores of 72.9 $\pm$ 3.3 and 71.3 $\pm$ 3.1 when tested with CE images and 73.4 $\pm$ 2.9 and 72.4 $\pm$ 3.1 when testing on non-CE images.

When using a model pre-trained with CE images (K$\_$CE$\_$Sampled or P$\_$CE$\_$Sampled) and fine tuning with CE images (K$\_$NoCE or P$\_$NoCE), Dice scores of 85.1 $\pm$ 2.1 and 84.9 $\pm$ 3.7 when testing with CE images were achieved. Dice scores of 81.3 $\pm$ 2.4 and 80.8 $\pm$ 3.1 were achieved when testing with non-CE images. 

Figure \ref{fig:mask_prost} shows some results for models trained using Mask R-CNN. Compared to nnU-Net, models pretrained with no CE data then fine tuned with CE data (c) and models trained with CE data (d) did not have outlying false positive predictions. However, the predicted masks were not as accurate compared to nnU-Net.

\subsection{Image Registration}

\begin{table}[ht]
\addtolength{\tabcolsep}{-2.5pt}
\scriptsize
\caption{Results from each of the experiments using VoxelMorph for image registration. The Dice score between the reference label and the moved label was recorded.} 
\label{tab:regresultsvox}
\begin{center}    
\begin{tabular}{|*{13}{c|}}  
\hline
\multicolumn{1}{|c}{} & \multicolumn{4}{|c|}{} & \multicolumn{8}{|c|}{Type of data used to train the model} \\ 
\cline{6-13}

\multicolumn{1}{|c}{} & \multicolumn{4}{|c}{} & \multicolumn{2}{|c}{} & 
\multicolumn{2}{|c}{} & \multicolumn{2}{|c}{} & \multicolumn{2}{|c|}{No CE} \\
\multicolumn{1}{|c}{Method} & \multicolumn{4}{|c}{Test Data} & \multicolumn{2}{|c}{Mixed} &
\multicolumn{2}{|c}{CE} & \multicolumn{2}{|c}{No CE} & \multicolumn{2}{|c|}{and early CE} \\
\multicolumn{1}{|c}{} & \multicolumn{4}{|c}{} & \multicolumn{2}{|c}{} &
\multicolumn{2}{|c}{} & \multicolumn{2}{|c}{} & \multicolumn{2}{|c|}{ uptake} \\
\cline{6-13} 

\multicolumn{1}{|c}{} & \multicolumn{4}{|c}{} & \multicolumn{1}{|c}{Kidney} & \multicolumn{1}{|c}{Prostate} &
\multicolumn{1}{|c}{Kidney} & \multicolumn{1}{|c}{Prostate} & \multicolumn{1}{|c}{Kidney} & \multicolumn{1}{|c}{Prostate} & \multicolumn{1}{|c}{Kidney} & \multicolumn{1}{|c|}{Prostate} \\\hline 

\multicolumn{1}{|c}{VoxelMorph} & \multicolumn{2}{|c}{\multirow{2}{*}{CE}} & \multicolumn{2}{|c}{\multirow{2}{*}{Dice $\pm$ std}} & \multicolumn{1}{|c}{\multirow{2}{*}{89.4 $\pm$ 4.1}} & \multicolumn{1}{|c}{\multirow{2}{*}{91.1 $\pm$ 3.8}} &
\multicolumn{1}{|c}{\multirow{2}{*}{\textbf{90.7 $\pm$ 3.2}}} & \multicolumn{1}{|c}{\multirow{2}{*}{\textbf{92.4 $\pm$ 2.6}}} & \multicolumn{1}{|c}{\multirow{2}{*}{71.4 $\pm$ 6.2}} & \multicolumn{1}{|c}{\multirow{2}{*}{76.8 $\pm$ 5.4}} & \multicolumn{1}{|c}{\multirow{2}{*}{78.6 $\pm$ 3.9}} & \multicolumn{1}{|c|}{\multirow{2}{*}{79.4 $\pm$ 3.8}} \\

\multicolumn{1}{|c}{(no} & \multicolumn{2}{|c}{} & \multicolumn{2}{|c}{} & \multicolumn{1}{|c}{} & \multicolumn{1}{|c}{} &
\multicolumn{1}{|c}{} & \multicolumn{1}{|c}{} & \multicolumn{1}{|c}{} & \multicolumn{1}{|c}{} & \multicolumn{1}{|c}{} & \multicolumn{1}{|c|}{}\\
\cline{4-13}

\cline{2-13}

\multicolumn{1}{|c}{pretraining)} & \multicolumn{2}{|c}{\multirow{2}{*}{No CE}} & \multicolumn{2}{|c}{\multirow{2}{*}{Dice $\pm$ std}} & \multicolumn{1}{|c}{\multirow{2}{*}{90.8 $\pm$ 3.4}} & \multicolumn{1}{|c}{\multirow{2}{*}{92.4 $\pm$ 2.8}} &
\multicolumn{1}{|c}{\multirow{2}{*}{75.8 $\pm$ 4.2}} & \multicolumn{1}{|c}{\multirow{2}{*}{79.4 $\pm$ 3.2}} & \multicolumn{1}{|c}{\multirow{2}{*}{\textbf{94.4 $\pm$ 2.3}}} & \multicolumn{1}{|c}{\multirow{2}{*}{\textbf{95.9 $\pm$ 2.2}}} & \multicolumn{1}{|c}{\multirow{2}{*}{84.6 $\pm$ 2.3}} & \multicolumn{1}{|c|}{\multirow{2}{*}{85.6 $\pm$ 2.3}} \\
\multicolumn{1}{|c}{} & \multicolumn{2}{|c}{} & \multicolumn{2}{|c}{} & \multicolumn{1}{|c}{} & \multicolumn{1}{|c}{} &
\multicolumn{1}{|c}{} & \multicolumn{1}{|c}{} & \multicolumn{1}{|c}{} & \multicolumn{1}{|c}{} & \multicolumn{1}{|c}{} & \multicolumn{1}{|c|}{}\\
\hline

\multicolumn{1}{|c}{VoxelMorph} & \multicolumn{2}{|c}{\multirow{2}{*}{CE}} & \multicolumn{2}{|c}{\multirow{2}{*}{Dice $\pm$ std}} & \multicolumn{1}{|c}{\multirow{2}{*}{-}} & \multicolumn{1}{|c}{\multirow{2}{*}{-}} &
\multicolumn{1}{|c}{\multirow{2}{*}{86.3 $\pm$ 4.3}} & \multicolumn{1}{|c}{\multirow{2}{*}{87.5 $\pm$ 4.2}} & \multicolumn{1}{|c}{\multirow{2}{*}{-}} & \multicolumn{1}{|c}{\multirow{2}{*}{-}} & \multicolumn{1}{|c}{\multirow{2}{*}{-}} & \multicolumn{1}{|c|}{\multirow{2}{*}{-}} \\
\multicolumn{1}{|c}{(pretrained} & \multicolumn{2}{|c}{} & \multicolumn{2}{|c}{} & \multicolumn{1}{|c}{} & \multicolumn{1}{|c}{} &
\multicolumn{1}{|c}{} & \multicolumn{1}{|c}{} & \multicolumn{1}{|c}{} & \multicolumn{1}{|c}{} & \multicolumn{1}{|c}{} & \multicolumn{1}{|c|}{}\\
\cline{4-13}
\cline{2-13}
\multicolumn{1}{|c}{with no} & \multicolumn{2}{|c}{\multirow{2}{*}{No CE}} & \multicolumn{2}{|c}{\multirow{2}{*}{Dice $\pm$ std}} & \multicolumn{1}{|c}{\multirow{2}{*}{-}} & \multicolumn{1}{|c}{\multirow{2}{*}{-}} &
\multicolumn{1}{|c}{\multirow{2}{*}{91.4 $\pm$ 2.8}} & \multicolumn{1}{|c}{\multirow{2}{*}{92.1 $\pm$ 2.9}} & \multicolumn{1}{|c}{\multirow{2}{*}{-}} & \multicolumn{1}{|c}{\multirow{2}{*}{-}} & \multicolumn{1}{|c}{\multirow{2}{*}{-}} & \multicolumn{1}{|c|}{\multirow{2}{*}{-}} \\
\multicolumn{1}{|c}{CE)} & \multicolumn{2}{|c}{} & \multicolumn{2}{|c}{} & \multicolumn{1}{|c}{} & \multicolumn{1}{|c}{} &
\multicolumn{1}{|c}{} & \multicolumn{1}{|c}{} & \multicolumn{1}{|c}{} & \multicolumn{1}{|c}{} & \multicolumn{1}{|c}{} & \multicolumn{1}{|c|}{}\\
\hline

\multicolumn{1}{|c}{VoxelMorph} & \multicolumn{2}{|c}{\multirow{2}{*}{CE}} & \multicolumn{2}{|c}{\multirow{2}{*}{Dice $\pm$ std}} & \multicolumn{1}{|c}{\multirow{2}{*}{-}} & \multicolumn{1}{|c}{\multirow{2}{*}{-}} &
\multicolumn{1}{|c}{\multirow{2}{*}{-}} & \multicolumn{1}{|c}{\multirow{2}{*}{-}} & \multicolumn{1}{|c}{\multirow{2}{*}{87.1 $\pm$ 1.6}} & \multicolumn{1}{|c}{\multirow{2}{*}{89.1 $\pm$ 1.5}} & \multicolumn{1}{|c}{\multirow{2}{*}{-}} & \multicolumn{1}{|c|}{\multirow{2}{*}{-}} \\
\multicolumn{1}{|c}{(pretrained} & \multicolumn{2}{|c}{} & \multicolumn{2}{|c}{} & \multicolumn{1}{|c}{} & \multicolumn{1}{|c}{} &
\multicolumn{1}{|c}{} & \multicolumn{1}{|c}{} & \multicolumn{1}{|c}{} & \multicolumn{1}{|c}{} & \multicolumn{1}{|c}{} & \multicolumn{1}{|c|}{}\\
\cline{4-13}
\cline{2-13}
\multicolumn{1}{|c}{with CE)} & \multicolumn{2}{|c}{\multirow{2}{*}{No CE}} & \multicolumn{2}{|c}{\multirow{2}{*}{Dice $\pm$ std}} & \multicolumn{1}{|c}{\multirow{2}{*}{-}} & \multicolumn{1}{|c}{\multirow{2}{*}{-}} &
\multicolumn{1}{|c}{\multirow{2}{*}{-}} & \multicolumn{1}{|c}{\multirow{2}{*}{-}} & \multicolumn{1}{|c}{\multirow{2}{*}{93.4 $\pm$ 2.4}} & \multicolumn{1}{|c}{\multirow{2}{*}{94.8 $\pm$ 2.3}} & \multicolumn{1}{|c}{\multirow{2}{*}{-}} & \multicolumn{1}{|c|}{\multirow{2}{*}{-}} \\
\multicolumn{1}{|c}{} & \multicolumn{2}{|c}{} & \multicolumn{2}{|c}{} & \multicolumn{1}{|c}{} & \multicolumn{1}{|c}{} &
\multicolumn{1}{|c}{} & \multicolumn{1}{|c}{} & \multicolumn{1}{|c}{} & \multicolumn{1}{|c}{} & \multicolumn{1}{|c}{} & \multicolumn{1}{|c|}{}\\
\hline

\end{tabular}
\end{center}
\end{table}

Table \ref{tab:regresultsvox} shows the results of the image registration using VoxelMorph. When testing each model with CE data, training a model using only CE data gave the best results. Training with K$\_$CE$\_$Sampled gave a Dice score of 90.7 $\pm$ 3.2 and training with P$\_$CE$\_$Sampled gave a Dice score of 92.4 $\pm$ 2.6. When testing on non-CE data, training a model using only non-CE data gave the best results. Training with K$\_$NoCE gave a Dice score of 94.4 $\pm$ 2.3 and training with P$\_$NoCE gave a Dice score of 95.9 $\pm$ 2.2.

A model pretrained with non-CE images (K$\_$NoCE or P$\_$NoCE) and fine tuned with CE images (K$\_$CE$\_$Sampled or P$\_$CE$\_$Sampled) achieved Dice scores of 86.3 $\pm$ 4.3 and 87.5 $\pm$ 4.2 when testing with CE images and 91.4 $\pm$ 2.8 and 92.1 $\pm$ 2.9 when testing with non-CE images. Alternatively, a model pretrained with sampled CE images (K$\_$CE$\_$Sampled or P$\_$CE$\_$Sampled) and fine tuned with non-CE images (K$\_$NoCE or P$\_$NoCE), achieved Dice scores of 87.1 $\pm$ 1.6 and 89.5 $\pm$ 1.5 when testing with CE images and 93.4 $\pm$ 2.4 and 94.8 $\pm$ 2.3 when testing with non-CE images.

\begin{figure*}[!htb]
\centering
\begin{minipage}[b]{.18\textwidth}
  \centering
  \centerline{\includegraphics[width=3cm]{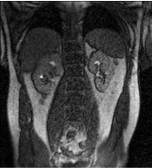}}
  \medskip
\end{minipage}
\begin{minipage}[b]{.18\textwidth}
  \centering
  \centerline{\includegraphics[width=3cm]{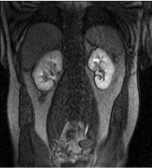}}
  \medskip
\end{minipage}
\begin{minipage}[b]{.18\textwidth}
  \centering
  \centerline{\includegraphics[width=3cm]{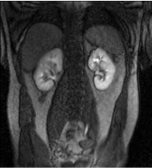}}
  \medskip
\end{minipage}
\begin{minipage}[b]{.18\textwidth}
  \centering
  \centerline{\includegraphics[width=3cm]{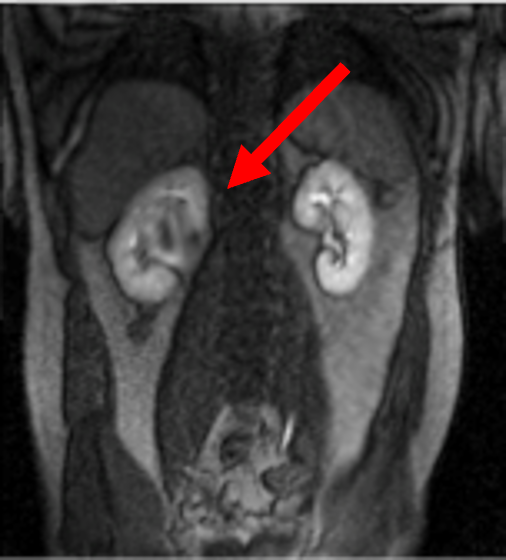}}
  \medskip
\end{minipage}
\begin{minipage}[b]{.18\textwidth}
  \centering
  \centerline{\includegraphics[width=3cm]{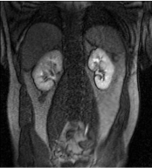}}
  \medskip
\end{minipage}

\begin{minipage}[b]{.18\textwidth}
  \centering
  \centerline{\includegraphics[width=3cm]{images/gt/kidneyce.png}}
  \centerline{(a) Reference}\medskip
\end{minipage}
\begin{minipage}[b]{.18\textwidth}
  \centering
  \centerline{\includegraphics[width=3cm]{images/gt/kidneynoce.png}}
  \centerline{(b) Moving}\medskip
\end{minipage}
\begin{minipage}[b]{.18\textwidth}
  \centering
  \centerline{\includegraphics[width=3cm]{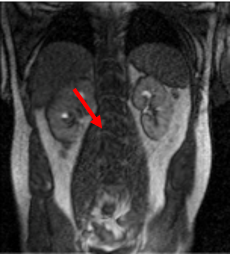}}
  \centerline{(c) Pretrained CE}\medskip
\end{minipage}
\begin{minipage}[b]{.18\textwidth}
  \centering
  \centerline{\includegraphics[width=3cm]{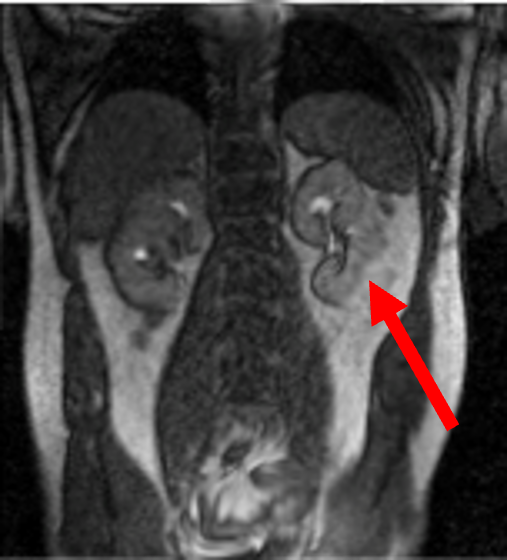}}
  \centerline{(d) No CE}\medskip
\end{minipage}
\begin{minipage}[b]{.18\textwidth}
  \centering
  \centerline{\includegraphics[width=3cm]{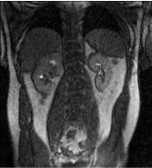}}
    \centerline{(e) CE}\medskip
\end{minipage}

\caption{VoxelMorph registration results on 2D kidney data. (a) reference image (top row is no CE and bottom row is CE), (b) moving image (top row is CE and bottom row is no CE), (c) model pretrained using CE and refined on non CE data, (d) model trained using no CE data, (e) model trained using CE data. Red arrows point towards examples of warps that have deformed the kidneys or spine.}
\label{fig:vox_kidney}
\bigskip
\medskip
\end{figure*}

Figure \ref{fig:vox_kidney} shows some results for models trained using VoxelMorph. Models trained using non CE data (d) gave results that did not generalise well. The red arrows point towards examples of warped kidneys.

\begin{table}[ht]
\addtolength{\tabcolsep}{-2.5pt}
\scriptsize
\caption{Results from each of the experiments using VTN for image registration. The Dice score between the reference label and the moved label was recorded.} 
\label{tab:regresultsvtn}
\begin{center}    
\begin{tabular}{|*{13}{c|}}  
\hline
\multicolumn{1}{|c}{} & \multicolumn{4}{|c|}{} & \multicolumn{8}{|c|}{Type of data used to train the model} \\ 
\cline{6-13}

\multicolumn{1}{|c}{} & \multicolumn{4}{|c}{} & \multicolumn{2}{|c}{} & 
\multicolumn{2}{|c}{} & \multicolumn{2}{|c}{} & \multicolumn{2}{|c|}{No CE} \\
\multicolumn{1}{|c}{Method} & \multicolumn{4}{|c}{Test Data} & \multicolumn{2}{|c}{Mixed} &
\multicolumn{2}{|c}{CE} & \multicolumn{2}{|c}{No CE} & \multicolumn{2}{|c|}{and early CE} \\
\multicolumn{1}{|c}{} & \multicolumn{4}{|c}{} & \multicolumn{2}{|c}{} &
\multicolumn{2}{|c}{} & \multicolumn{2}{|c}{} & \multicolumn{2}{|c|}{ uptake} \\
\cline{6-13} 

\multicolumn{1}{|c}{} & \multicolumn{4}{|c}{} & \multicolumn{1}{|c}{Kidney} & \multicolumn{1}{|c}{Prostate} &
\multicolumn{1}{|c}{Kidney} & \multicolumn{1}{|c}{Prostate} & \multicolumn{1}{|c}{Kidney} & \multicolumn{1}{|c}{Prostate} & \multicolumn{1}{|c}{Kidney} & \multicolumn{1}{|c|}{Prostate} \\\hline 

\multicolumn{1}{|c}{VTN} & \multicolumn{2}{|c}{\multirow{2}{*}{CE}} & \multicolumn{2}{|c}{\multirow{2}{*}{Dice $\pm$ std}} & \multicolumn{1}{|c}{\multirow{2}{*}{88.4 $\pm$ 3.4}} & \multicolumn{1}{|c}{\multirow{2}{*}{86.1 $\pm$ 4.9}} &
\multicolumn{1}{|c}{\multirow{2}{*}{\textbf{90.1 $\pm$ 3.1}}} & \multicolumn{1}{|c}{\multirow{2}{*}{\textbf{89.5 $\pm$ 3.8}}} & \multicolumn{1}{|c}{\multirow{2}{*}{70.1 $\pm$ 5.9}} & \multicolumn{1}{|c}{\multirow{2}{*}{68.4 $\pm$ 7.2}} & \multicolumn{1}{|c}{\multirow{2}{*}{75.3 $\pm$ 4.2}} & \multicolumn{1}{|c|}{\multirow{2}{*}{74.1 $\pm$ 4.1}} \\

\multicolumn{1}{|c}{(no} & \multicolumn{2}{|c}{} & \multicolumn{2}{|c}{} & \multicolumn{1}{|c}{} & \multicolumn{1}{|c}{} &
\multicolumn{1}{|c}{} & \multicolumn{1}{|c}{} & \multicolumn{1}{|c}{} & \multicolumn{1}{|c}{} & \multicolumn{1}{|c}{} & \multicolumn{1}{|c|}{}\\
\cline{4-13}

\cline{2-13}

\multicolumn{1}{|c}{pretraining)} & \multicolumn{2}{|c}{\multirow{2}{*}{No CE}} & \multicolumn{2}{|c}{\multirow{2}{*}{Dice $\pm$ std}} & \multicolumn{1}{|c}{\multirow{2}{*}{89.3 $\pm$ 2.8}} & \multicolumn{1}{|c}{\multirow{2}{*}{87.3 $\pm$ 4.2}} &
\multicolumn{1}{|c}{\multirow{2}{*}{74.3 $\pm$ 3.4}} & \multicolumn{1}{|c}{\multirow{2}{*}{73.1 $\pm$ 4.5}} & \multicolumn{1}{|c}{\multirow{2}{*}{\textbf{92.4 $\pm$ 2.1}}} & \multicolumn{1}{|c}{\multirow{2}{*}{\textbf{91.1 $\pm$ 3.4}}} & \multicolumn{1}{|c}{\multirow{2}{*}{82.4 $\pm$ 2.8}} & \multicolumn{1}{|c|}{\multirow{2}{*}{81.2 $\pm$ 3.8}} \\
\multicolumn{1}{|c}{} & \multicolumn{2}{|c}{} & \multicolumn{2}{|c}{} & \multicolumn{1}{|c}{} & \multicolumn{1}{|c}{} &
\multicolumn{1}{|c}{} & \multicolumn{1}{|c}{} & \multicolumn{1}{|c}{} & \multicolumn{1}{|c}{} & \multicolumn{1}{|c}{} & \multicolumn{1}{|c|}{}\\
\hline

\multicolumn{1}{|c}{VTN} & \multicolumn{2}{|c}{\multirow{2}{*}{CE}} & \multicolumn{2}{|c}{\multirow{2}{*}{Dice $\pm$ std}} & \multicolumn{1}{|c}{\multirow{2}{*}{-}} & \multicolumn{1}{|c}{\multirow{2}{*}{-}} &
\multicolumn{1}{|c}{\multirow{2}{*}{85.1 $\pm$ 3.8}} & \multicolumn{1}{|c}{\multirow{2}{*}{83.4 $\pm$ 3.7}} & \multicolumn{1}{|c}{\multirow{2}{*}{-}} & \multicolumn{1}{|c}{\multirow{2}{*}{-}} & \multicolumn{1}{|c}{\multirow{2}{*}{-}} & \multicolumn{1}{|c|}{\multirow{2}{*}{-}} \\
\multicolumn{1}{|c}{(pretrained} & \multicolumn{2}{|c}{} & \multicolumn{2}{|c}{} & \multicolumn{1}{|c}{} & \multicolumn{1}{|c}{} &
\multicolumn{1}{|c}{} & \multicolumn{1}{|c}{} & \multicolumn{1}{|c}{} & \multicolumn{1}{|c}{} & \multicolumn{1}{|c}{} & \multicolumn{1}{|c|}{}\\
\cline{4-13}
\cline{2-13}
\multicolumn{1}{|c}{with no} & \multicolumn{2}{|c}{\multirow{2}{*}{No CE}} & \multicolumn{2}{|c}{\multirow{2}{*}{Dice $\pm$ std}} & \multicolumn{1}{|c}{\multirow{2}{*}{-}} & \multicolumn{1}{|c}{\multirow{2}{*}{-}} &
\multicolumn{1}{|c}{\multirow{2}{*}{89.9 $\pm$ 3.2}} & \multicolumn{1}{|c}{\multirow{2}{*}{87.4 $\pm$ 3.9}} & \multicolumn{1}{|c}{\multirow{2}{*}{-}} & \multicolumn{1}{|c}{\multirow{2}{*}{-}} & \multicolumn{1}{|c}{\multirow{2}{*}{-}} & \multicolumn{1}{|c|}{\multirow{2}{*}{-}} \\
\multicolumn{1}{|c}{CE)} & \multicolumn{2}{|c}{} & \multicolumn{2}{|c}{} & \multicolumn{1}{|c}{} & \multicolumn{1}{|c}{} &
\multicolumn{1}{|c}{} & \multicolumn{1}{|c}{} & \multicolumn{1}{|c}{} & \multicolumn{1}{|c}{} & \multicolumn{1}{|c}{} & \multicolumn{1}{|c|}{}\\
\hline

\multicolumn{1}{|c}{VTN} & \multicolumn{2}{|c}{\multirow{2}{*}{CE}} & \multicolumn{2}{|c}{\multirow{2}{*}{Dice $\pm$ std}} & \multicolumn{1}{|c}{\multirow{2}{*}{-}} & \multicolumn{1}{|c}{\multirow{2}{*}{-}} &
\multicolumn{1}{|c}{\multirow{2}{*}{-}} & \multicolumn{1}{|c}{\multirow{2}{*}{-}} & \multicolumn{1}{|c}{\multirow{2}{*}{89.3 $\pm$ 2.4}} & \multicolumn{1}{|c}{\multirow{2}{*}{88.5 $\pm$ 3.1}} & \multicolumn{1}{|c}{\multirow{2}{*}{-}} & \multicolumn{1}{|c|}{\multirow{2}{*}{-}} \\
\multicolumn{1}{|c}{(pretrained} & \multicolumn{2}{|c}{} & \multicolumn{2}{|c}{} & \multicolumn{1}{|c}{} & \multicolumn{1}{|c}{} &
\multicolumn{1}{|c}{} & \multicolumn{1}{|c}{} & \multicolumn{1}{|c}{} & \multicolumn{1}{|c}{} & \multicolumn{1}{|c}{} & \multicolumn{1}{|c|}{}\\
\cline{4-13}
\cline{2-13}
\multicolumn{1}{|c}{with CE)} & \multicolumn{2}{|c}{\multirow{2}{*}{No CE}} & \multicolumn{2}{|c}{\multirow{2}{*}{Dice $\pm$ std}} & \multicolumn{1}{|c}{\multirow{2}{*}{-}} & \multicolumn{1}{|c}{\multirow{2}{*}{-}} &
\multicolumn{1}{|c}{\multirow{2}{*}{-}} & \multicolumn{1}{|c}{\multirow{2}{*}{-}} & \multicolumn{1}{|c}{\multirow{2}{*}{91.3 $\pm$ 2.9}} & \multicolumn{1}{|c}{\multirow{2}{*}{90.7 $\pm$ 3.5}} & \multicolumn{1}{|c}{\multirow{2}{*}{-}} & \multicolumn{1}{|c|}{\multirow{2}{*}{-}} \\
\multicolumn{1}{|c}{} & \multicolumn{2}{|c}{} & \multicolumn{2}{|c}{} & \multicolumn{1}{|c}{} & \multicolumn{1}{|c}{} &
\multicolumn{1}{|c}{} & \multicolumn{1}{|c}{} & \multicolumn{1}{|c}{} & \multicolumn{1}{|c}{} & \multicolumn{1}{|c}{} & \multicolumn{1}{|c|}{}\\
\hline

\end{tabular}
\end{center}
\end{table}

Table \ref{tab:regresultsvtn} shows the results of the image registration tasks using VTN. When testing each model with CE data, training a model using only CE data gave the best results. Training with K$\_$CE$\_$Sampled gave a Dice score of 90.1 $\pm$ 3.1 and training with P$\_$CE$\_$Sampled gave a Dice score of 89.5 $\pm$ 3.8. When testing on non-CE data, training a model using only non-CE data gave the best results. Training with K$\_$NoCE gave a Dice score of 92.4 $\pm$ 2.1 and training with P$\_$NoCE gave a Dice score of 91.1 $\pm$ 3.4.

A model pretrained with non-CE images (K$\_$NoCE or P$\_$NoCE) and fine tuned with CE images (K$\_$CE$\_$Sampled or P$\_$CE$\_$Sampled) achieved Dice scores of 85.1 $\pm$ 3.8 and 83.4 $\pm$ 3.7 when testing with CE images and 89.9 $\pm$ 3.2 and 87.4 $\pm$ 3.9 when testing with non-CE images. Alternatively, a model pretrained with sampled CE images (K$\_$CE$\_$Sampled or P$\_$CE$\_$Sampled) and fine tuned with sampled non-CE images (K$\_$NoCE or P$\_$NoCE) achieved Dice scores of 89.3 $\pm$ 2.4 and 88.5 $\pm$ 3.1 when testing with CE images and 91.3 $\pm$ 2.9 and 90.7 $\pm$ 3.5 when testing with non-CE images.

\begin{figure*}[!htb]
\centering
\begin{minipage}[b]{.18\textwidth}
  \centering
  \centerline{\includegraphics[width=3cm]{images/gt/kidneynoce.png}}
  \medskip
\end{minipage}
\begin{minipage}[b]{.18\textwidth}
  \centering
  \centerline{\includegraphics[width=3cm]{images/gt/kidneyce.png}}
  \medskip
\end{minipage}
\begin{minipage}[b]{.18\textwidth}
  \centering
  \centerline{\includegraphics[width=3cm]{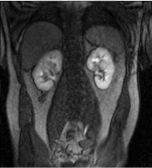}}
  \medskip
\end{minipage}
\begin{minipage}[b]{.18\textwidth}
  \centering
  \centerline{\includegraphics[width=3cm]{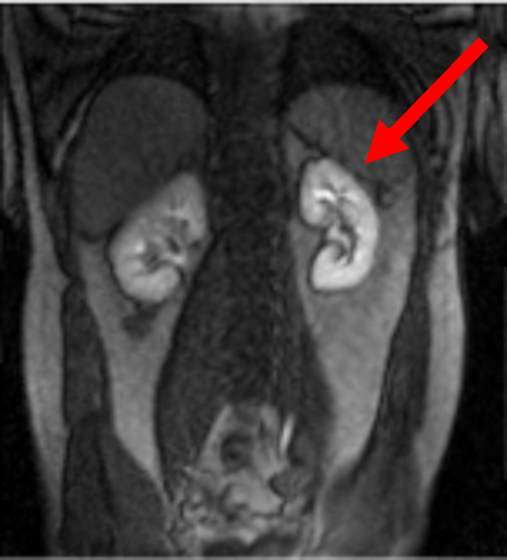}}
  \medskip
\end{minipage}
\begin{minipage}[b]{.18\textwidth}
  \centering
  \centerline{\includegraphics[width=3cm]{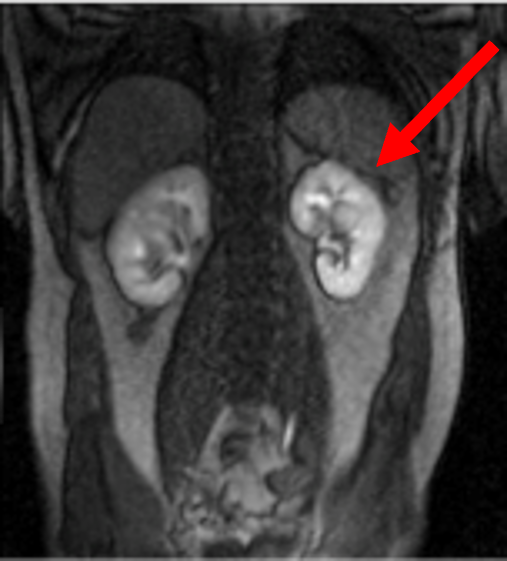}}
  \medskip
\end{minipage}

\begin{minipage}[b]{.18\textwidth}
  \centering
  \centerline{\includegraphics[width=3cm]{images/gt/kidneyce.png}}
  \centerline{(a) Reference}\medskip
\end{minipage}
\begin{minipage}[b]{.18\textwidth}
  \centering
  \centerline{\includegraphics[width=3cm]{images/gt/kidneynoce.png}}
  \centerline{(b) Moving}\medskip
\end{minipage}
\begin{minipage}[b]{.18\textwidth}
  \centering
  \centerline{\includegraphics[width=3cm]{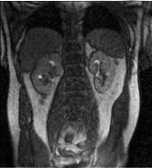}}
  \centerline{(c) Pretrained CE}\medskip
\end{minipage}
\begin{minipage}[b]{.18\textwidth}
  \centering
  \centerline{\includegraphics[width=3cm]{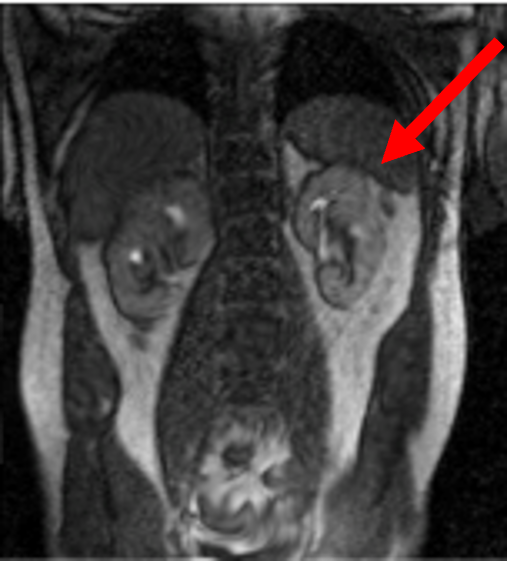}}
  \centerline{(d) No CE}\medskip
\end{minipage}
\begin{minipage}[b]{.18\textwidth}
  \centering
  \centerline{\includegraphics[width=3cm]{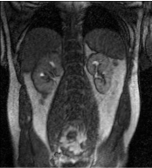}}
    \centerline{(e) CE}\medskip
\end{minipage}

\caption{VTN registration results on 2D kidney data. (a) reference image (top row is no CE and bottom row is CE), (b) moving image (top row is CE and bottom row is no CE), (c) model pretrained using CE and refined on non CE data, (d) model trained using no CE data, (e) model trained using CE data. Red arrows point towards examples of warps that have deformed the kidneys.}

\label{fig:vtn_kidney}
\end{figure*}

Figure \ref{fig:vtn_kidney} shows some results for models trained using VTN. Models trained using non CE data (d) gave results that did not generalise well. For models trained with CE data (e), when registering a moving image with CE to a reference image with no CE the kidney is warped to a larger shape than in both original images. The red arrows point towards examples of warped kidneys.

\section{DISCUSSION}
In this work, we explored different architectures for both segmentation and image registration so that the effects of CE can be explored. 

For segmentation, training using datasets containing a mixture of contrast and non CE images provided the best result when testing with CE images, while training using CE data and fine tuning with non-CE data gave the best result when testing with non-CE data. Moreover, the latter only exhibited a small decrease in performance compared to training with mixed data when testing with CE data. This is interesting as it outperformed pre-training a model with non-CE data and fine tuning with CE data. Noteworthy, the CE dataset contains data which has very little CE up to the maximum amount of CE leading to a much more diverse dataset than the non-CE data. Pre-training the model using the CE data may have helped the model to learn more diverse low level features allowing for better generalisable models. Moreover, these findings suggest that it is important to have as much diversity as possible in the early stages of training for both nnU-Net and Mask R-CNN.

For the image registration task, training and testing with the same type of data gave the best results. This is most likely due to the similarity metric that is used during training. One of the key issues with image registration of DCE-MRI is finding an image similarity metric that can measure the alignment whilst having intensity differences. Compared to other common similarity metrics, MI is the most suitable. Indeed, it can work well with multi-modal data (such as T1w and T2w MRI) as it is tolerant of intensity differences between images if they are structurally close and these differences are consistent. However, changes in CE can make this a challenging task in DCE-MR images. Pre-training using CE data and then fine tuning using non-CE data gave models that generalised well. Indeed, only a slight decrease in performance for testing on CE data was observed but an increase when testing with non-CE data. The standard deviation was also much smaller.

We found that strategically using the available data by splitting according to CE instead of using the whole dataset gave models that generalised better for each task and architecture. This increase in performance gained from pre-training using CE data may give the model examples of data that are easier to learn from as well as having a diverse set of data due to the different amounts of CE. The ROI in images without CE can often be difficult to distinguish from surrounding tissue, therefore, we believe that thanks to a better edge definition in the ROI due to CE, the data is then easier to learn from. Learning low level features from these easier examples can thus reduce the challenge of learning the same features from harder examples. This is further supported by the results of the models trained using the peak and non CE datasets. For each task and architecture, models that were trained using non CE data have the lowest performance in each test case, except for image registration of non CE data. When CE data was used to train the models, we saw a global increase in performance. Interestingly, while the peak dataset has images with and without CE, training with the mixed dataset produced better results. A strong hypothesis would be that, while the peak dataset contains CE images, its distribution is heavily skewed towards data without CE or with only a small amount of CE making it harder to learn than the CE dataset. Pretraining with CE data likely performed better than models pretrained with non CE data due to learning better weights from the CE data allowing for better feature maps to be predicted from the non CE data. Interestingly, these findings closely follow the principles of curriculum learning \cite{bengio09}; using the data in a meaningful order from easiest to most difficult examples.

\section{CONCLUSION}
In this paper we challenge the common assumption that contrast enhancement inevitably increases the difficulty of deep learning based image processing tasks. Our evaluation of deep learning methods for segmentation (nnU-Net and Mask R-CNN) and registration (VoxelMorph and VTN) found that using a more diverse dataset earlier in the training process can lead to more generalisable models.

\section*{ACKNOWLEDGMENTS} 
This work was funded by Medical Research Scotland and Canon Medical Research Europe.


\bibliography{report} 

\begin{thebibliography}{10}

\bibitem{Kershaw2006}
L.~E. Kershaw and D.~L. Buckley, ``Precision in measurements of perfusion and
  microvascular permeability with t1-weighted dynamic contrast-enhanced mri,''
  {\em Magnetic Resonance in Medicine}, vol.~56, 2006.

\bibitem{Mahapatra2015}
D.~Mahapatra, Z.~Li, F.~Vos, and J.~Buhmann, ``Joint segmentation and groupwise
  registration of cardiac dce mri using sparse data representations,'' {\em
  IEEE 12th International Symposium on Biomedical Imaging (ISBI)}, 2015.

\bibitem{Warfield2018}
M.~Haghighi, S.~K. Warfield, and S.~Kurugol, ``Automatic renal segmentation in
  dce-mri using convolutional neural networks,'' {\em IEEE 15th International
  Symposium on Biomedical Imaging (ISBI)}, 2018.

\bibitem{Lietzmann2012}
F.~Lietzmann, F.~G. Zöllner, U.~I. Attenberger, S.~Haneder, H.~J. Michaely,
  and L.~R. Schad, ``Dce-mri of the human kidney using blade: A feasibility
  study in healthy volunteers,'' {\em Journal of Magnetic Resonance Imaging},
  vol.~35, 2012.

\bibitem{lemaitre}
G.~Lemaître, R.~Martí, J.~Freixenet, J.~C. Vilanova, P.~M. Walker, and
  F.~Meriaudeau, ``Computer-aided detection and diagnosis for prostate cancer
  based on mono and multi-parametric mri: A review,'' {\em Computers in Biology
  and Medicine}, vol.~60, 2015.

\bibitem{Isensee2021}
F.~Isensee, P.~F. Jaeger, S.~A. Kohl, J.~Petersen, and K.~H. Maier-Hein,
  ``nnu-net: a self-configuring method for deep learning-based biomedical image
  segmentation,'' {\em Nature Methods}, vol.~18, 2021.

\bibitem{He2017}
K.~He, G.~Gkioxari, P.~Dollar, and R.~Girshick, ``{Mask R-CNN},'' {\em
  Proceedings of the IEEE International Conference on Computer Vision}, 2017.

\bibitem{Ren2017}
S.~Ren, K.~He, R.~Girshick, and J.~Sun, ``{Faster R-CNN: Towards Real-Time
  Object Detection with Region Proposal Networks},'' {\em IEEE Transactions on
  Pattern Analysis and Machine Intelligence}, vol.~39, 2017.

\bibitem{He2016}
K.~He, X.~Zhang, S.~Ren, and J.~Sun, ``{Deep residual learning for image
  recognition},'' {\em Proceedings of the IEEE Computer Society Conference on
  Computer Vision and Pattern Recognition}, 2016.

\bibitem{yi2016}
T.~Lin, P.~Doll{\'{a}}r, R.~B. Girshick, K.~He, B.~Hariharan, and S.~J.
  Belongie, ``Feature pyramid networks for object detection,'' {\em Computing
  Research Repository (CoRR)}, 2016.

\bibitem{Balakrishnan2019}
G.~Balakrishnan, A.~Zhao, M.~R. Sabuncu, J.~Guttag, and A.~V. Dalca,
  ``Voxelmorph: A learning framework for deformable medical image
  registration,'' {\em IEEE Transactions on Medical Imaging}, vol.~38, 2019.

\bibitem{Zhao2019}
S.~Zhao, T.~Lau, J.~Luo, E.~I.-C. Chang, and Y.~Xu, ``Unsupervised 3d
  end-to-end medical image registration with volume tweening network,'' {\em
  {IEEE} Journal of Biomedical and Health Informatics}, vol.~24, 2020.

\bibitem{bengio09}
Y.~Bengio, J.~Louradour, R.~Collobert, and J.~Weston, ``Curriculum learning,''
  ICML '09, Association for Computing Machinery, 2009.

\end{thebibliography}
\bibliographystyle{ieeetr} 

\end{document}